# MODELING OF H- ION SOURCE AT LANSCE*

Nikolai Yampolsky†, Ilija Draganic, and Lawrence Rybarcyk,
Los Alamos National Laboratory, Los Alamos, USA.

## Abstract

We report on the progress in modelling performance of the H- ion source at LANSCE. The key aspect we address is the lifetime of the tungsten filament. The lifetime depends on multiple parameters of the ion source and can dramatically vary in different regimes of operation. We use the multiphysics approach to model the performance of the ion source. The detailed analysis has been made to recognize key physical processes, which affect the degradation of the filament. The analysis resulted in the analytical model, which includes relevant processes from the first principles. The numerical code based on this model has been developed and benchmarked. The results of the modelling show good agreement with experimental data. As a result, the developed model allows predicting the performance of the ion source in various regimes of operation.

## INTRODUCTION

The Los Alamos Neutron Science Center (LANSCE) [1] hosts an 800 MeV linear accelerator, which delivers H+ and H- beams to multiple experimental facilities including the Isotope Production Facility (IPF), Lujan Center (neutron spallation source), the Weapon Neutron Research (WNR), the Proton Radiography Facility (pRAD), and the Ultra Cold Neutron (UCN) experiments. The H- beam is produced be the multicusp surface converter ion source [2, 3]. The lifetime of the ion source is limited by the lifetime of the tungsten filaments, which are used to produce arc and create plasma. Earlier research resulted in models capable of predicting the ballpark of the filament lifetime [4, 5]. However, those models include several empirical factors, introduced to match the experimental observations. It is an indication that earlier models miss key physics and they are not likely to describe significantly different regimes of the ion source operation.

In this paper we describe the progress in development of modeling capabilities of the H- ion source at LANL. We develop a multiphysics model describing the evolution of the hot filaments from the first principles. We have identified the major physical effects which affect the parameters of the filament and include them into the model.

## FILAMENT MODEL

The filament is heated by the current distributed along the filament. The filament is not heated uniformly since the current along the filament is not constant due to presence of arc current. The non-uniform distribution of temperature affects local parameters of the filament material, such as: resistivity, thermal conductivity, evaporation rate, emissivity, thermionic emission, *etc*. These parameters, in turn, define the distribution of current. Such a physics requires a self-consistent model for the filament parameters.

The filament is described as a wire of round cross section. The diameter of the wire, as well as its temperature, vary along the filament. The filament is assumed to have no variation in the cross section.

### Ohm's Law

The distribution of current along the filament is described by the Ohm's law. The current is driven by the direct current (DC) voltage applied to the filament ends.

$$\frac{dU}{dz} = I\frac{dR}{dz}, \quad (1)$$

$$\frac{dR}{dz} = \frac{\rho(z)}{\pi d^2(z)/4}, \quad (2)$$

$$\frac{dI}{dz} = (j_e + j_i)\pi d(z), \quad (3)$$

where $U(z)$ is the distribution of electrostatic potential along the filament, $I(z)$ is the distribution of current along the filament, $dR/dz$ is the differential resistance, $\rho(z)$ is the material resistivity which depends on local temperature, $d(z)$ is local diameter of the filament, and $j_e$ and $j_i$ are the electron and ion arc current densities, respectively. Equations (1) – (3) should be solved with the boundary condition of $U(L)-U(0)=U_{DC}$, where $L$ is the filament length and $U_{DC}$ is the applied DC voltage. Equations (1) – (3) should be solved separately for phases of the ion source cycle with and without the arc current. The electron and ion arc current densities can be found from Richardson's law for thermionic emission [6] and plasma sheath problem [7], respectively. The schematics of current flows through the filament is shown in Fig. 1.

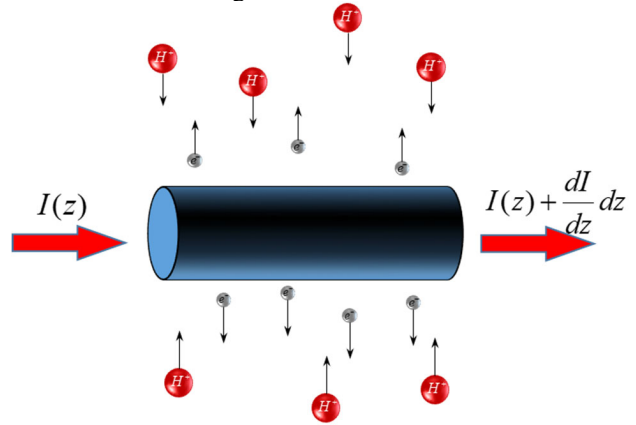

Figure 1: Schematics of currents flow through filament.

### Energy Balance Equation

Local properties of the material (*e.g.* electron emission and resistivity) depend on the temperature distribution along the filament. As a result, Ohm's law should be complimented with the energy balance equation, which defines the distribution of temperature along the filament. This equation balances the inflow and the outflow of power to the filament

$$\frac{d}{dz}\left(\kappa \frac{\pi d^2}{4}\frac{dT}{dz}\right) - \varepsilon\sigma(T^4 - T_{env}^4) + \frac{\pi d^2}{4}\sum \dot{q}_l = 0, \quad (4)$$

where $T(z)$ is the temperature distribution along the filament, $T_{env}$ is the temperature of the environment, i.e. temperature of the ion source walls, $\kappa$ is the thermal conductivity, $\sigma \approx 5.67 \cdot 10^{-8}$ W/(m²·K⁴) is the Stephan-Boltzmann constant, $\varepsilon$ is the material emissivity, and $\dot{q}_l$ are the volumetric power inflows to the filament from multiple physical processes. The first term in Eq. (1) describes thermal conductivity along the filament. It is dominant at the filament ends and ensures smooth transition between cold filament ends (which match temperature of the poles) and the hot central part. The second term in Eq. (4) describes the filament cooling due to the emission of the black body radiation. This is the main mechanism for stabilizing the temperature in the middle part of the filament. The last term in Eq. (4) describes the power inflow to the filament due to different physical processes. The individual physical processes which are accounted for by the model and the corresponding power inflows are listed in Table 1.

---

* Work supported by the US Department of Energy under Contract Number DE-AC52-06NA25396
† Email address: nyampols@lanl.gov

Table 1: Contribution from Various Processes

| Process | Volumetric power flow |
|---|---|
| Ohmic heating | $[(1-f)I^2 + fI_{arc}^2]\frac{dR}{dz}$ |
| Ion heating | $fj_i\pi d(\phi_{arc} - U)$ |
| Electron cooling | $-fj_e\pi d\phi_W$ |
| Cooling by neutrals | $\sqrt{\frac{9}{2\pi}\frac{T_{gas}}{m_{gas}}}\frac{4n_{gas}}{d}(T_{gas} - T)$ |

Table 1 has the following notations: $f$ is the ion source duty factor, $I_{arc}$ is the distribution of current along the filament during the arc phase of operation while $I$ is the current distribution when arc voltage is not applied, $\phi_{arc}$ is the applied arc voltage, $\phi_W$ is the work function of the material, $T_{gas}$ is the temperature of the surrounding gas, $n_{gas}$ is the gas density, and $m_{gas}$ is the mass of neutral molecules of the gas.

Ohmic heating is the main mechanism for heating the filament. The current distribution is different when the arc voltage is applied or turned off and filament is heated slightly differently during these phases of the cycle. As a result, the filament is slightly hotter on the side of negative potential. Eventually, the filament degrades faster on that side. This effect was first observed and explained in [8]. Our estimates show that the ion source repetition rate of 120 Hz is large enough so that the filament temperature does not change during different phases of operation.

Ion heating is caused by the energetic plasma ions bombarding the filament during arc phase of operation. The ions are accelerated by the sheath and reach energies close to the applied arc potential. Similarly, the electron arc current results in the filament cooling since each electron needs to transition a potential barrier of the work function in order to be emitted into plasma. That energy is taken away from the stored thermal energy of the filament bulk material.

The cooling by neutrals is caused by the thermal energy exchange between the filament and the surrounding gas. It is assumed that neutrals are colliding with the filament having the average thermal energy of the gas and bounce back having the thermal energy of the filament. The molecules of the gas are assumed to have Gaussian distribution, which defines the flux of particles and average deposited energy. The temperature of the gas is estimated through the energy balance of the gas. The arc deposits energy to the gas and it is extracted from the ion source chamber through the wall mediated by neutrals-wall collisions. The temperature distribution of the gas is assumed be to uniform inside the source chamber.

The contributions to the power balance equation from individual physical mechanisms are summarized in Table 2. The numbers in the Table 2 are approximate and demonstrate the importance of each individual effect rather than provide a precise value. The estimates are made based on the simulation which models ion source at LANSCE.

Table 2: Contribution from Individual Mechanisms

| Process | Power [Watts] |
|---|---|
| Radiation losses | -1000 |
| DC Ohmic heating | 1100 |
| Ohmic heating by arc current | 40 |
| Thermal conductivity | -80 |
| Ion heating | 20 |
| Electron cooling | -5 |
| Cooling by neutrals | -15 |

*Degradation of Filament*

There are two effects which cause degradation of the filament. The diameter of the filament reduces over time due to evaporation of the material. The filament is degraded faster at high temperature. That limits the lifetime of the filament if the ion source is operated at high arc current (and correspondingly at high ion current). Additionally, the filament is degraded due to sputtering, which is caused by bombardment of the filament with the energetic cesium ions. The Cs ions are accelerated by the plasma sheath and have the energy identical to the energy of the hydrogen ions, which contribute to ion heating. The reduction of the filament diameter over time is described by the following equation

$$\frac{\partial}{\partial t}d = -2\frac{C_E}{D} - 2Yj_i\delta_{Cs}\frac{m_{Cs}}{-eD}, \quad (5)$$

where $C_E$ is the evaporation rate of the filament material, $D$ is is the filament density, $\delta_{Cs}$ is the fraction of Cs ions in plasma, $Y = Y(\phi_{arc} - U)$ is the yield of filament sputtering by Cs ions [9], $m_{Cs}$ is the mass of Cs ions, $e$ is the electron charge.

## INPUT AND OUT PARAMETERS

The developed model is self-consistent. The diameter and the temperature profiles of the filament are rigorously calculated from the first principles and no empirical assumptions are made to fit data. The model requires several parameters as an input. First, the model starts with the uniform filament of a given diameter. The model is extremely sensitive to the initial diameter (about 1.55 mm), and it has to be indicated with $10^{-3}$ accuracy in order to match experimental data. Next, the parameters of plasma need to be indicated. Currently, we do not self-consistently simulate plasma parameters. Plasma density of $\sim 10^{12}$ cm$^{-3}$ and electron temperature of 1.5 eV provide reasonable agreement between modeling and experimental data and these values are consistent with plasma parameters of other ion sources [10]. The fraction of Cs ions in plasma is also an external parameter and is on the order of $\delta_{CS} \sim 10^{-4}$ for typical run cycles. Gas density corresponds to pressure of 3 mTorr at room temperature. The temperature of filament poles and the ion source chamber is 60 °C. The applied DC voltage and arc voltage are provided into the model from EPICS.

Once the initial parameters are chosen, the diameter and temperature profiles of the filament are calculated from the first principles. The model generates the time history for the filament resistance and the arc current. These are the parameters which are used for comparison with experimental data.

## MODELING LANSCE H- ION SOURCE

We have used the developed model to simulate degradation of the filaments in H- ion source at LANSCE. Each run cycle has been operated at slightly different conditions (repetition rate of 60 Hz and 120 Hz, $\phi_{arc} \sim 150\text{-}170$ V, $U_{DC}$=11-12 V, and the resulting arc current of 25-45 A provided by two filaments). Comparison between the model and the experimental data for a typical cycle is shown in Fig. 2.

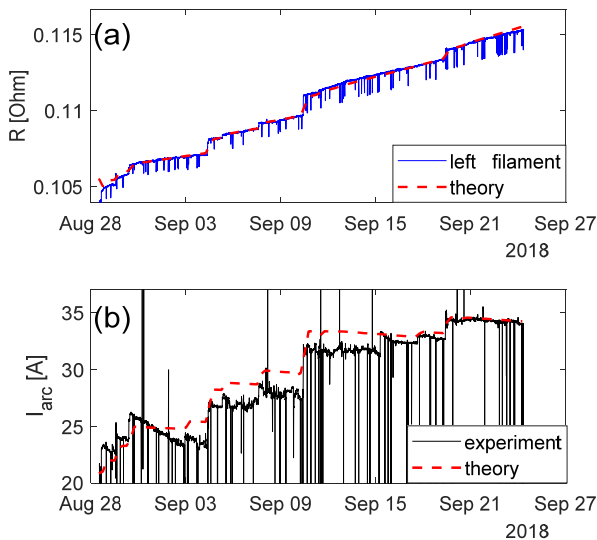

Figure 2: Comparison of results produced by model (red dashed lines) with experimental data (solid lines).

In general, the model shows good agreement with the experimental data once the initial parameters are carefully adjusted. The model accurately describes the change in the filament resistance (Fig. 2a) and arc current (Fig. 2b). The changes in arc current are reasonably depicted by the model while the regime of the ion source operation significantly changes over time.

The initial parameters of the model which explain the dynamics of the filament were consistent across different cycles. For example, the plasma density was different within a factor of 2 between different runs and was typically higher at higher arc current. There are several mechanisms which may contribute to the remaining discrepancies between the modeling and the experiment, as well as inconsistencies between initial conditions of different run cycles. First, the model may still lack some essential physics, which has not been included. Second, the diagnostics may have some systematic errors. Finally, the performance of the negative ion source is sensitive to the preparation and operation processes. Some effects, like inconsistent vacuum or outgassing may cause significant variations between different run cycles.

## CONCLUSION

We have developed a numerical model which self-consistently describes the operation of filaments in gas discharge. This model has been benchmarked against experimental data of H- ion source delivering beam at LANSCE. The model shows good agreement with experimental data. That allows one to use such a model to estimate the life time of the filament in various regimes of operation.

## ACKNOWLEDGEMENTS

Authors are thankful to Sergey Kurrennoy and David Kleinjan for useful discussions.

## REFERENCES

[1] LANSCE, https://lansce.lanl.gov/

[2] R. L. York, *et al.*, "The development of a high current H− injector for the proton storage ring at LAMPF", *Nucl. Instrum. Meth. Phys. Res. Sect. B*, vol. 10-11, p. 891, 1985.

[3] K. W. Ehlers and K. N. Leung, "Multicusp negative ion source," *Rev. Sci. Instrum.*, vol. 51, p.721, no. 6, 1980.

[4] E. Chacon-Golcher, "Erosion and failure of tungsten filament," LANL, Los Alamos, USA, Rep. LA-UR-08–5251, 2008.

[5] I. N. Draganic, J. F. O'Hara, and L. J. Rybarcyk, "Different approaches to modeling the LANSCE H− ion source filament performance", *Rev. Sci. Instrum.*, vol. 87, p. 02B112, 2015.

[6] O. W. Richardson "On the negative radiation from hot platinum", *Proc. Cambridge Phil. Soc.*, vol. 11 p. 286, 1901.

[7] I. Langmuir, "Positive ion currents from the positive column of Mercury arcs," *Science,* vol. 58, p. 290, 1923.

[8] I. N. Draganic, J. F. O, and L. Rybarcyk, "Lifetime Study of Tungsten Filaments in an H- Surface Convertor Ion Source", in *Proc. North American Particle Accelerator Conf. (NAPAC'13)*, Pasadena, CA, USA, Sep.-Oct. 2013, paper THPAC23, pp. 1190-1192.

[9] Institut für Angewandte Physik, http://www.iap.tuwien.ac.at/www/

[10] K. Jayamanna, et al., "A 60 mA DC H- multi cusp ion source developed at TRIUMF", *Nucl. Instrum. Meth. Phys. Research A*, vol. 895, p. 150, 2018.